\documentstyle[11pt,dunk2001_asp,twoside,psfig]{article}
\def\edcomment#1{\iffalse\marginpar{\raggedright\sl#1\/}\else\relax\fi}
\reversemarginpar

\begin{document}
\title{Extragalactic Planetary Nebulae}
 \author{Holland Ford and Eric Peng}
\affil{Johns Hopkins University, Homewood Campus, Baltimore, MD 21013, USA}
\author{Ken Freeman}
\affil{Research School of Astronomy \& Astrophysics, Australian National
 University, Canberra, ACT 2611}

\begin{abstract}
Planetary nebulae (PNe) can be used to trace intermediate and old stellar
populations in galaxies and the intracluster medium out to
approximately 20 Mpc.  PNe can be easily identified with
narrow-band surveys, and their chemical abundances and radial velocities
can be measured with multi-slit or multi-fiber spectrographs.  We briefly review
what has been accomplished to date, and include tables (with references) that
summarize the observations of PNe in 42 galaxies and in the intracluster media of the
Virgo and Fornax clusters. The surveys for intracluster PNe suggest that between
20$\%$ and 50$\%$ of the clusters' total stellar mass is in the intracluster medium!
We illustrate the utility of
PNe for dynamical studies by presenting results from a
recently completed survey of Cen A (NGC~5128) with the NOAO mosaic
camera on the CTIO 4-m telescope. The PNe and globular cluster surveys 
extend respectively to distances of 80 Kpc and 50 Kpc.  We compare the kinematics
of the stars in Cen A (736 PN velocities) to the
 kinematics of the red and 
blue globular clusters (188 clusters with
measured velocities).  The latter 
include 125 newly confirmed globular
clusters from our survey.
\end{abstract}

\section{Introduction}
Planetary nebulae (PNe) can be used to trace old and intermediate-age
stellar populations in galaxies and in the intracluster medium out to
approximately 20 Mpc. Photometric and spectroscopic observations of PNe
provide information about: 1) the stellar death rate, 2) the chemical 
composition of the planetaries' parent stars, 3) the distance to the galaxy,
and 4) the stellar dynamics of the parent stellar population and the
distribution of mass in the galaxy.

Planetary nebulae in the intracluster or intragroup medium (ICPNe)
provide a measure of past tidal stripping and tidal disruption of
galaxies within the cluster or group.  Depending on the evolutionary
state of the cluster, ICPNe may provide {\it thousands} of test particles that
can be used to study the cluster's dynamics.

\section{Identification of PNe and Camera/Telescopes Suitable for PNe Surveys}

Planetary nebulae mark the endpoint of stellar evolution for stars with
masses between $\sim0.8$ and $8$ M$_{\sun}$ (Iben \& Renzini, 1983; Dopita et
al. 1997).  The expanding shell (or shells) of a planetary nebula is
ejected during the ascent of the asymptotic giant branch.  The shell is
ionized by the ultraviolet radiation from the newly exposed, hot
helium-burning core.  A large fraction of the radiation from the core,
which has a luminosity equal to a red giant, is absorbed via ionization
of hydrogen and helium, and converted into emission lines from the most
abundant elements.  The temperature of the nebula is set by equilibrium
between heating from ionization and cooling from permitted and forbidden
emission lines. Radiation from [OIII]~$\lambda4959/5007$ is one of the most
efficient cooling lines; the monochromatic luminosity in
[OIII]~$\lambda5007$ is comparable to the V-band luminosity of the
K-giants in old stellar populations. Consequently, PNe can be
efficiently identified by taking pairs of on-band and off-band
[OIII]~$\lambda5007$ images. When using ground-based telescopes, PNe in galaxies 
beyond the Magellanic Clouds will be unresolved, absent in the off-band image, 
and brighter in [OIII]~$\lambda5007$ than H$\alpha$.  Detailed descriptions of PNe
survey techniques and issues that must be considered when selecting and
buying interference filters can be found in Ford et al.\ (1989) and Jacoby et
al. (1992). 

Telescope and camera combinations that are particularly suitable for PNe
surveys are summarized in Table~1.  Column five gives
the product of the telescope's aperture times the field of view, A$\times\Omega$,
normalized to the Subaru telescope's ``Suprime" camera. All else being
equal (seeing and camera throughput), A$\times\Omega$  is a figure of
merit for a camera/telescope's survey capability. The last column gives
the distance at which a PN one magnitude fainter than the brightest PN will
have a signal-to-noise ratio of 10 in the sum of four 2250 second exposures.
We assumed that the seeing is 0.75$\arcsec$ FWHM, the net atmosphere/telescope/camera 
throughput is 0.5, the readnoise is 6 e$^{-}$ RMS per pixel, and the sky brightness
and isophotal galaxy brightness are $V = 22$ mag/$\Box\arcsec$.  For the MMT we assumed
that the sky and galaxy surface brightnesses are 21.8, and that the CCDs are 
binned $3 \times 3$ before readout.  

\begin{table}
\caption{Telescopes and Cameras  That Are Suitable for PNe Surveys}
\begin{tabular}{cccccc}
\tableline

Telescope	& Camera  	& Pixel Size	& FOV    	&  A$\times\Omega$	& Dist\cr
		& Format	& ($\arcsec$)	& (arc-min)	& 		& (Mpc) \cr	
\tableline
\tableline

WHT 4.2-m/PNS	& 2k $\times$ 2k	& 0.33 $\times$ 0.30	& 11.3 $\times$ 10.3	& 0.045 & 20 \cr
ESO 2.2-m	& 8k $\times$8k	& 0.24	& 34 $\times$ 33	& 0.12 &14 \cr
CTIO 4-m	& 8k $\times$ 8k	& 0.27	& 36 $\times$ 36	& 0.45 & 20\cr
MMT 6.5-m	& 18k $\times$ 18k	& 0.08	& 24 $\times$ 24	& 0.66 & 25\cr
CFHT 3.6-m	& 20k $\times$ 18k	& 0.18	& 62 $\times$ 56	& 0.97 & 18 \cr
Subaru 8.2-m	& 10k $\times$ 18k	& 0.20	& 30 $\times$ 24	& 1 & 29 \cr
\tableline
\end{tabular}
\end{table}

The Planetary Nebula Spectrograph (PNS) (Arnaboldi et al.\ 2001; Douglas,
2001) on the William Herschel Telescope is a slitless spectrograph
preceded by a narrow band [OIII]~$\lambda5007$ filter.  The spectrograph is
rotated $180^{\deg}$ between exposures.  When the images are compared, PNe will
be pairs of stellar sources whose separations are proportional to their
radial velocities, whereas stars will be dispersed into short
spectra.  The PNS will be very effective for measuring the radial velocities of
PNe for two reasons.  When the seeing is very good, as frequently is the
case on the WHT (Wilson et al.\ 1999), the effective slit will be the
seeing FWHM.  And, a second observing run is not required to measure the
radial velocities.

The Hubble Advanced Camera for Surveys will be able to detect ICPNe one
magnitude fainter than the brightest PN at a distance of $\sim200$ Mpc when
the sky is as faint or fainter than 23.2 V-mag/$\Box\arcsec$.

\section{Abundance Determinations Using PNe}
Spectroscopic observations of planetary nebulae are presently the only
way to measure the chemical abundances of individual elements in old and
intermediate age stars at distances of $\sim1$ Mpc or greater.
Although the nebular abundances of He, C, N, and S are enhanced by
nucleosynthesis in the parent star (e.g. Dopita et al.\ 1997), comparisons
can be made of relative abundances from one part of a galaxy to another,
and between galaxies.

In the LMC, [O/H] in bright PNe has the same value as [O/H] in HII
regions (Richer 1993); consequently, the oxygen abundance measured in
PNe is a good measure of the oxygen abundance in the gas from which the
parent stars formed. 

With the exception of the He/H ratio, direct abundance determinations
require the measurement of the electron temperature $T_e$, the electron
density $n_e$, and line intensities in two or more ionization
stages. Because the crucial temperature diagnostic line
[OIII]~$\lambda4363$ is 50 to 200 times weaker than [OIII]~
$\lambda5007$, the
measurement of this faint line sets the distance limit for abundance
determinations.  To date, direct abundance measurements have not been
made in galaxies more distant than M31.  Indirect PNe abundance
determinations based on diagnostic line ratios have been made in
NGC~5128 (Walsh et al.\ 1999).

Jacoby \& Ciardullo (1999) measured the chemical abundances in 15 M31
PNe.  Their paper, which finds a wide range in [O/H] in M31's old stellar populations,
is a model for these  difficult observations.

\section{The Planetary Nebula Luminosity Function}

Image intensifier detections of PNe in M31 and its dwarf companions 
(NGC~147, NGC~185, NGC~205, and M32) with the Lick 3-m telescope (Ford,
Jenner, \& Epps 1973; Ford \& Jenner 1975; Ford, Jacoby, \& Jenner 1977;
Ford \& Jacoby 1978 ) and subsequent Image Tube Scanner spectrophotometry
showed that the brightest nebulae had approximately the same
[OIII]~$\lambda5007$ flux.  We reasoned that if we compared similar old stellar
populations, the main sequence turn-off masses would be similar.
Consequently, there would be a maximum number of ionizing photons that
could be produced by a planetary's central star, and a maximum
[OIII]~$\lambda5007$ flux.  Using this reasoning, we proposed that
planetary nebulae might be used as standard candles (Ford \& Jenner, 1978).  Although the
utility of the planetary nebula luminosity function (PNLF) has been
controversial (Sandage \& Tammann 1990; Bottinelli 1991;Tammann 1993), a
great deal of observational work (e.g. Ciardullo et al.\ 1989a; Ciardullo
et al.\ 1989b; Jacoby et al. 1989) showed that the PNLF in
the bulges of spiral galaxies and in early type galaxies is an excellent
standard candle with little or no dependence on Hubble type.
Analytical, theoretical, and numerical models (Jacoby, 1989; Dopita,
Jacoby, \& Vassiliadis 1992; Mendez 1993) established the theoretical
basis for the observational fact that the PNLF is a good standard
candle.  These studies showed that the PNLF has only a modest dependence
on the ages and metallicities of the parent population, and the number
of PNe observed.  However, the {\it observed} dependencies appear to be much
smaller than those predicted by the models.  Jacoby, Walker, and
Ciardullo (1990) found that the observed PNLF distance to the LMC is in
excellent agreement with the LMC Cepheid distance.

Ciardullo et al.\ (1989a; C89a) developed the model for the PNLF that has
been used so successfully for deriving distances to galaxies that do not
have cepheids.  Their representation of the PNLF is given in equation~1.

\begin{equation}
N(M) \propto e^{0.307M} (1-e^{3(M^{*}-M)})
\end{equation}

$M^{*}$ in equation 1 cuts off the luminosity function and serves as a
``standard candle". Based on a distance of 770 Kpc to M31, the calibration of the PNLF was
given by $M^{*} = -4.48$ (C89a). Using Madore \& Freedman's (1991)
Cepheid distance to M31, Ciardullo et al.\ (1998) derived 
$M^{*} = -4.54$. More recently Ferrarese et al.\ (2000) derived 
$M^{*} = -4.58$ by comparing PNLF distances with HST Cepheid distances to 
the same galaxies, or to galaxies in the same group or cluster. They found a very 
tight linear relationship
between the PNLF and Cepheid distances. The PNLF 
distances to the Virgo and
Fornax clusters appear to be systematically smaller than the Cepheid
distances.  This may be due to intracluster PNe; in a magnitude-limited
sample, these will be preferentially on the nearside of the cluster (see Section 6).

The relationship 
between the apparent monochromatic magnitude $m_{5007}$ 
and the observed monochromatic flux $F_{5007}$ is given by

\begin{equation} 
m_{5007} = -2.5{\rm log} F_{5007} - 13.74. 
\end{equation}

The [OIII]~$\lambda5007$ flux from an $M^{*}$ planetary at 10 parsecs is $1.977\times10^{-4}$ ergs cm$^{-2}$ s$^{-1}$.

\section{Summary of Surveys for Extragalactic PNe}

At the time of writing, approximately 5000 PNe have been identified in
more than 40 external galaxies. These observations are summarized in
Table~2. The galaxy types were taken from the RSA (Sandage \& Tammann 1980, 1987) 
when available.  Galaxy distances in bold type are based on the PNLF. Column 4 gives 
the number of PNe identified in each galaxy and column 5 gives the references for the 
identifications.   The
sixth column list references for those galaxies that have chemical 
abundance determinations
for their PNe.  Space did not permit inclusion of the large number 
of chemical abundance studies of PNe in the LMC and SMC.  References for galaxies with PNe 
radial velocities and dynamical studies are listed in the last column of Table~2.

\begin{table}
\caption{ Observations of Extragalactic Planetary Nebulae}
 \begin{tabular}{llllllll}
\tableline                                                                                                                                                 
Galaxy	&	Type	&	Dist	&	No. 	&	ID	&	Chem	&	Radial	\cr
	&		&	(Mpc)	&	of PNe	&	Refs	&	Abund	&	Velocity	\cr
	&		&		&		&		&	Refs	& References		\cr
\tableline													
\tableline													
													
Sag Df	&	ImIV-V	&	0.025	&	2	&	Z96	&	W97	&	Z96	\cr
Fornax	&	Irr III-IV	&	& 1	&	D78	&	D78,W97	&	D78	\cr	
SMC	&	ImIV-V	&	{\bf 0.062}	&	62	&	M95	&	*	&F68	\cr	
LMC	&	Irr	&	{\bf 0.049}	&	 280	&	L97	&	*	&F68, S72	\cr	
LMC	&		&		&		&		&		&	F79b,M88	\cr
N 147	&	dE5	&		&	5	&	F77	&		&	F77	\cr
N 185	&	dE3p	&	0.57	&	5	&	F73,F77	&	R95	&	F77	\cr
N 205	&	S0/E5p	&	0.76	&	28	&	F73,C89b	&	R95	&		\cr
N 6822	&	ImIV-V	&		&	7	&	K82	&	K82,R95	&		\cr
M31	&	SbI-II	&	0.77	&	1000	&	F78, L82	&	J86,S98,	&	L82, L83	\cr
M31	&		&		&		&	C89b	& J99,R99 	&	N87, H94	\cr	
M31	&		&		&		&		&	H00 	&	E00	\cr
M32	&	E2	&	$< 0.77$	&	30	&	F73, F75	&	J79,S98	&	N86	\cr
M32	&		&		&		&	C89b	&	R99,H00	&		\cr
M33	&	Sc(s)II-III	&	0.84	&	138	&	M01, C01	&	 	&		\cr
M33	&		&{\bf 0.85}	&	58	&	K99	&		&		\cr	
IC 10	&	dIrr	&		&	1	&	J81	&		&		\cr
Leo A	&	dIrr	&		&	2	&	J81	&		&		\cr
Sex A	&	dIrr	&		&	1	&	J81	&	 	&		\cr
Peg	&	dI/dSph	&		&	1	&	J81	&		&		\cr
WLM	&	IrrIV-V	&		&	2	&	J81	&	 	&		\cr
N 300	&	ScII.8	&	{\bf 2.4}	&	34	&	S96	&	 	&		\cr
M81	&	Sb(r)I-II	&	{\bf 3.50}	&	185	&	J89	&	 	&		\cr
N 2403	&	Sc(s)III	&	{\bf 3.26}	&	40	&	K99	&		&		\cr
N 5128	&	S0+S pec	&	{\bf 3.5}	&	1140	&	H93b,P02	&	W99	&	H93a,H95	\cr
N 5128	&		&		&		&	F02	&		&	M96, M99	\cr
N 5128	&		&		&		&		&		&	P02	\cr
N 5102	&	S0$\_$1(5)	&{\bf 3.1}	&	26	&	M94a	&	 	&		\cr	
N 4736	&	SA(r)ab	&	6.0	&	67	&	D00	&		&	D00	\cr
N 3627 	&	Sb(s)II.2	&	{\bf 10.41}	&	73	&	K99	&		&		\cr
N 3377	&	E6	&	{\bf 10.3}	&	54	&	C89a	&		&	 	\cr
N 3379	&	E0	&{\bf 9.8}	&	93	&	C89a	&	 	&	C93, S99	\cr	
N 3379	&		&		&		&		&		&	S00	\cr
N 3384	&	SB0	&	{\bf 10.1}	&	102	&	C89a	&		&	 T95	\cr
N 3368	&	 Sab(s)II	&	{\bf 9.6}	&	74	&	F96B	&		&		\cr
N 1023	&	SB0$\_$1(5)	&	{\bf 9.86}	&	110	&	C91	&		&		\cr
N 891	&	Sb 	&	{\bf 9.86}	&	33	&	C91	&		&		\cr
N 5194	&	Sbc(s)I-II	&{\bf 8.4}	&	64	&	F96b	&		&		\cr	
N 5457	&	Sc(s)I	&{\bf 7.7}	&	65	&	F96a,F96b	&		&		\cr	
N 4594	&	Sa+/Sb-	&	{\bf 8.9}	&	294	&	F96c	&		&		\cr
N 4374	&	E1	&	{\bf 15.7}	&	102	&	J90	&		&		\cr
N 4382	&	S0$\_$1(3) pec	&	{\bf 14.4}	&	102	&	J90	&		&		\cr
N 4406	&	S0$\_$1(3)/E3	&	{\bf 15.7}	&	141	&	J90	&		&	 A96	\cr
N 4472	&	E1/S0$\_$1(1)	&	{\bf 13.9}	&	54	&	J90	&		&		\cr
\tableline
\end{tabular}
\end{table}

\setcounter{table}{1}
\begin{table}
\caption{  (Continued)}
 \begin{tabular}{llllllll}
\tableline                                                                                                                                                 
Galaxy	&	Type	&	Dist	&	No. 	&	ID	&	Chem	&	RV	\cr
	&		&	(Mpc)	&	of Pne	&	Refs	&	Abund	&	Refs	\cr
	&		&		&		&		&	Refs	&		\cr
\tableline													
\tableline			
N 4486	&	E0	&	{\bf 13.3}	&	338	&	J90, C98	&	 	&		\cr
N 4649	&	S0$\_$1(2)	&	{\bf 14.2}	&	32	&	J90	&		&		\cr
N 4478	&	E2	&	{\it 13.3}	&	7	&	C98	&		&		\cr
IC3443	&	dE0	&		&	1	&	C98	&		&		\cr
N 1316	&	Sa pec	&	{\bf 16.8}	&	105	&	M94b	&		&	A98	\cr
N 1399	&	E1	&	{\bf 17.1}	&	72	&	M94b	&		&	S00	\cr
N 1404	&	E2	&	{\bf 17.0}	&	47	&	M94b	&		&		\cr
\tableline
\end{tabular}
\end{table}

\section{Intracluster Planetary Nebulae}
The discovery of intracluster planetary nebulae (ICPNe) in the Virgo and
Fornax clusters is one of the most interesting new developments in the
study of extragalactic PNe.  ICPNe were first discovered by Arnaboldi et
al. (1996; A96) in a survey of the Virgo galaxy NGC~4406. Spectra were
taken of PNe candidates (Jacoby et al.\ 1990; J90) in two fields that
were centered 134\arcsec E and 134\arcsec W of NGC~4406.  Sixteen of the PNe
candidates in the two fields have radial velocities near N4406's
systemic velocity of $-227$ km~s$^{-1}$.  Three of the candidates in the W
field have velocities of 1729 km~s$^{-1}$, 1651 km~s$^{-1}$, and 1340 km~s$^{-1}$, 
provided the observed emission line is [OIII]~$\lambda5007$.  At these redshifts, 
the [OIII]~$\lambda4959$ emission line falls in J90's on-band filter 
($\lambda_c\sim4998$~\AA, FWHM$\sim30$~\AA), accounting for the detection of 
the high velocity PNe.
Because the nebulae were identified with a filter centered at 4998~\AA,
the emission lines detected spectroscopically at $\sim 5033$~\AA~ {\it cannot} be Ly$\alpha$ 
from galaxies with redshifts $z \sim3.14$.  A96 suggested that these three nebulae are
PNe from a Virgo intracluster stellar population.

Subsequent surveys of the Virgo and Fornax clusters have identified ~300
emission line sources that may be ICPNe.  The surveys are summarized in
Table~3. Column 5 lists the survey authors' estimated fraction of the cluster's 
total stellar mass that is in the intracluster medium. These estimates suggest
that 20$\%$ to 50$\%$ of a cluster's stellar mass may be in the intracluster medium!  

Depending on the depth of the survey, 25\% or more of the PNe
candidates may be starbursting galaxies at $z \sim 3.1$ with Ly$\alpha$ emission
redshifted into the [OIII]~$\lambda5007$ on-band filter (cf Kudritzki et
al. 2000; Krelove et al. 1999; Freeman et al. 2000).   Provided there is a 
sufficient signal-to-noise ratio in the
spectra, PNe can be distinguished from starbursting galaxies and QSOs at
$z\sim3.1$ by the presence of [OIII]~$\lambda4959$ and the absence of a
continuum.
At spectral resolutions of $\sim10$~\AA\, or higher, the Ly$\alpha$ 
emission line will be
resolved, whereas [OIII]~$\lambda5007$ in a PN with an expansion velocity
of $\sim20$ km~s$^{-1}$ will be unresolved.

\begin{table}
\caption {ICPNe Candidates in the Virgo and Fornax Clusters}
\begin{tabular}{cccccc}
\tableline
Cluster	& Fields 	& Survey Area 	& Num. of ICPNe & IC-Stellar &Refs \cr	
	&	&(sq-arcmin)	&{\it Candidates}& Fraction		&  \cr
\tableline
\tableline	
Virgo	& N 4406 	& 32	& 3 	& $-$	& A96 \cr 
Virgo	& Virgo Core 	& 50 	& 11 	& 0.5	&  M87 \cr 
Virgo	& 2 Intracluster 	& 512 	 & 85 	& 0.2	& F98 \cr
Virgo	& M87 Halo 	& 256 	& $\sim 75$ 	& $-$	& C98 \cr
Virgo	& Intracluster 	& $-$	& 23 confirmed	& $-$	& F00 \cr
Fornax	&  Intra-cluster 	& 104	& 10	& 0.4	& T97 \cr
Fornax	& Intra-cluster 	& 0.58 sq-deg	& $\sim135$	& 0.15 - 0.2	& K00 \cr
\tableline
\end{tabular}
\end{table}

Several arguments suggest that a large fraction of the ICPNe candidates
are in fact intracluster planetary nebulae.  Freeman et al.\ (2001)
spectroscopically confirmed 23 ICPNe in the Virgo cluster by detecting
[OIII]~$\lambda4959$ and [OIII]~$\lambda5007$ 
at the expected wavelengths and
intensity ratio.  The anomalous PNLF in M87's halo has PNe that are
brighter than the PNe in M87's main body, and thus are likely
foreground ICPNe (Ciardullo et al.\ 1998).  SN 1980I occurred midway
between NGC~4374 and NGC~4406,
showing that there are intracluster stars in the Virgo cluster (Smith,
1981).  Ferguson, Tanvir, \& von Hippel  (1998) used HST images to detect
faint intracluster stars in isolated Virgo fields. This population of
(old) stars will produce planetary nebulae. 

ICPNe are important for many reasons.  They may provide {\it thousands} of
test particles for detailed studies of the mass distribution in
clusters. Their kinematics may reveal otherwise unobservable tidal
streams that record tidal interactions over the last few Giga-years. 

Little is known about the presence of intragroup stars in small groups
of galaxies. 
If groups of galaxies have the same fraction of luminous intergalactic
material as estimated for Virgo and Fornax, there could be hundreds of
intergalactic PNe.  The kinematics of these nebulae could be used to
investigate the distribution of mass in sparse groups, and to study the
history of tidal interactions. 

\section{Stellar Populations in Cen A's Halo}
PNe and globular clusters (GCs) are both used as probes of galaxy halos.
While PNe sample the field star population, and have emission lines that
are more conducive to identification and radial velocity measurements,
GCs are old star clusters for which we also can obtain metallicity and
size information.  Together, PNe and GCs are complementary tools for
studying the structure and evolutionary histories of early-type
galaxies.

At a distance of 3.5 Mpc (Hui et al.\ 1993; H93b), Centaurus A (NGC~5128) is
the nearest massive elliptical galaxy, making it an excellent target for
PNe and GC studies.  Previous PNe surveys (H93b, Hui et al.\ 1995; H95)
identified 785 PNe out to 20 Kpc along the major axis and 10 Kpc along the
minor axis, and
resulted in velocities for 433 PNe.  We recently completed an extended
survey that reaches projected radii of 80 and 40 Kpc along the major and
minor axes, respectively.  This brings the total number of PNe to 1140,
of which 736 have velocities.  In addition, we conducted a UBVRI survey
with spectroscopic follow-up out to 50 and 30 Kpc to find new GCs and
study the faint halo light.

\begin{figure}
\vskip -2.25cm
\centerline{\vbox{
\psfig{figure=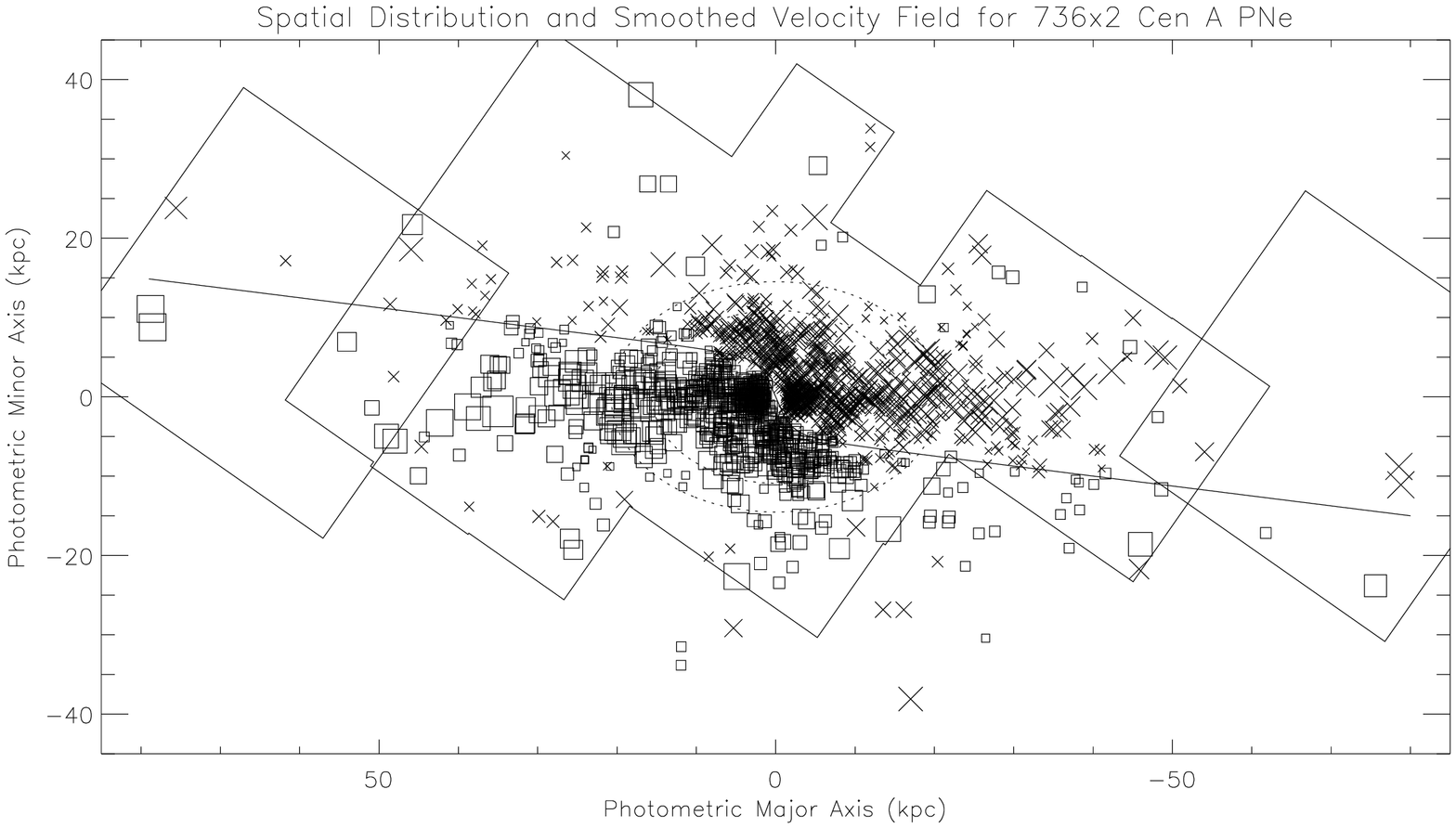,width=13cm,angle=90}
\centerline{\hbox{
\psfig{figure=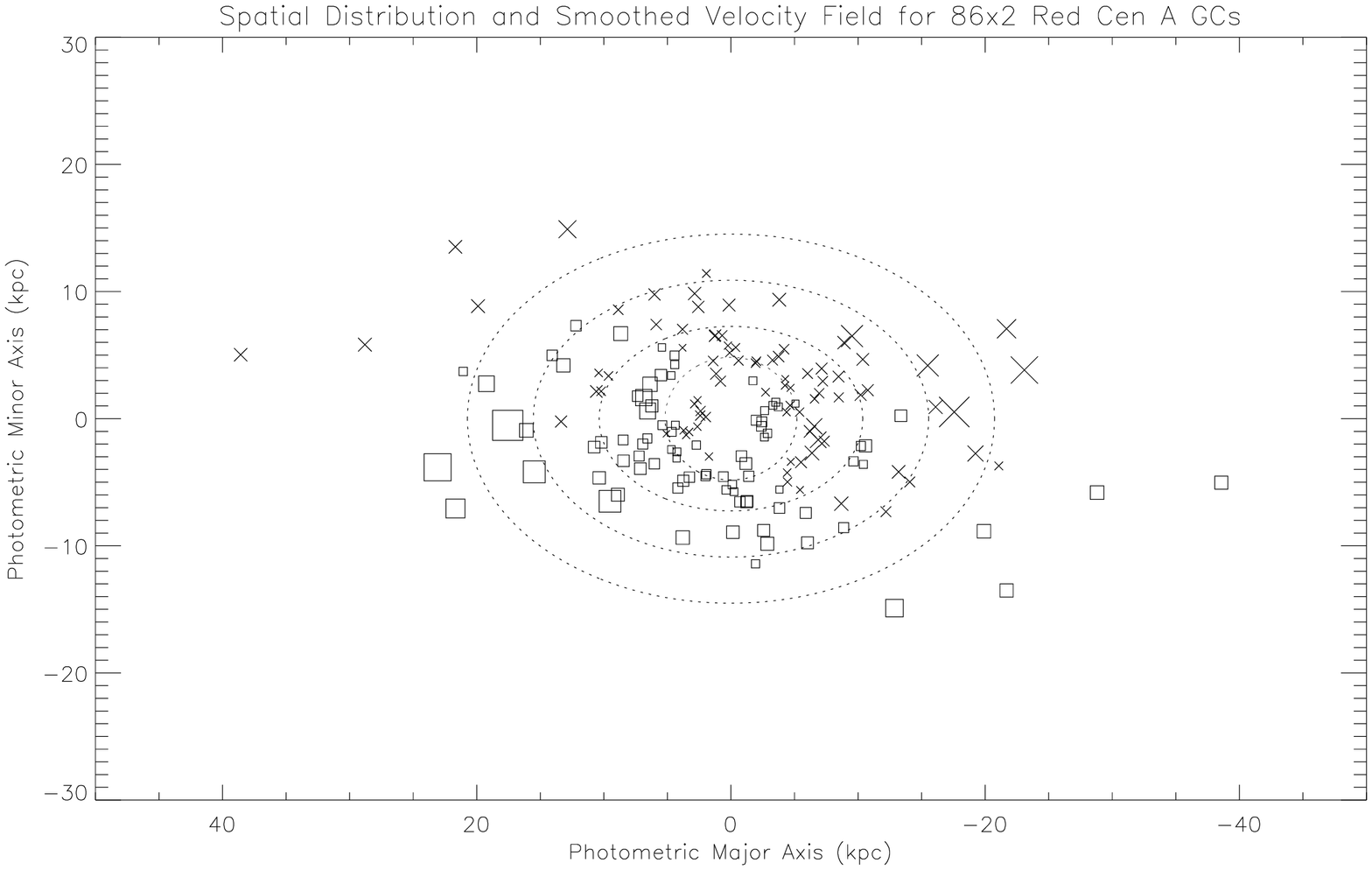,width=6.5cm,angle=90}
\psfig{figure=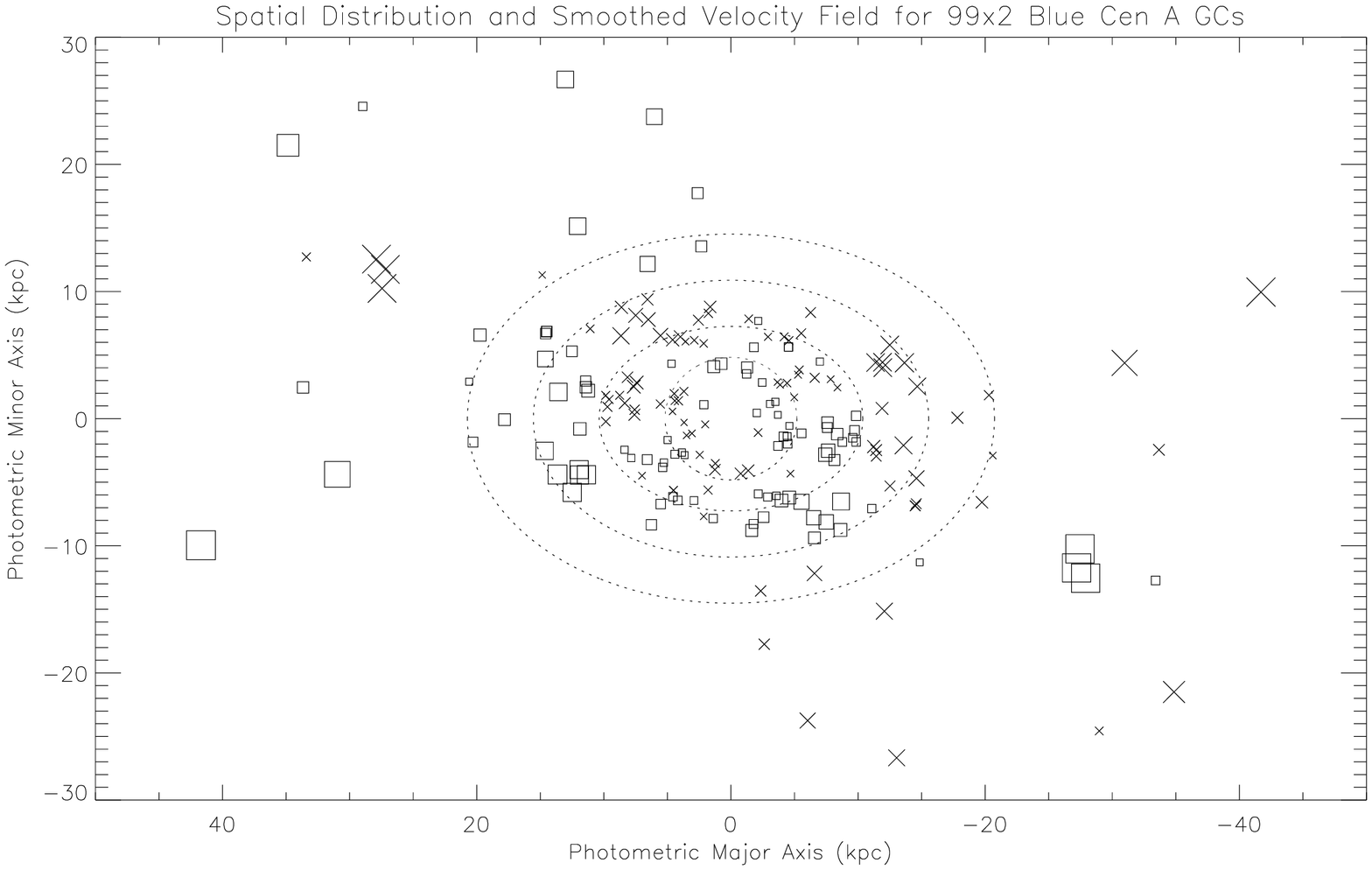,width=6.5cm,angle=90}
}}}}
\caption{{\em Cen A PN and GC Smoothed Velocity Fields.}  
The x$/$y-axes are the photometric major$/$minor axes.
The dotted 
ellipses are 1--4 $r_e$ isophotes for the old
stellar light.  Each point is a
confirmed PN or GC. The mean 
velocity field at
each object is the radial velocity average of
all objects within $r \le 3$ Kpc.  In sparse regions, the
velocities are averaged with nearby neighbors.
Velocities greater$/$less 
than systemic are represented by an ``x''$/$box, with
a point size proportional to the absolute deviation from systemic
(up to $\sim\pm150$ km/s).  We assume
point symmetry (valid in a triaxial potential), and reflect each object through 
the origin while reversing the sign of its velocity wrt the galaxy. a) (top) 
{\em The PN velocity field.}  Overplotted are the outlines of our survey 
(large boxes), and the
zero velocity contour.  Points appear to exist outside our survey
area because of the assumed point symmetry,
creating a 1472 particle field.  The zero-velocity contour has a strong
twist in the inner regions.  b) (bottom left) {\em The Red GC Velocity
Field.}  There are 86 GCs redder than ($V$--$I$)$_0=0.98$, the dip
in the bimodal color distribution.  These GCs are more centrally
concentrated than the blue GCs, and rotate in a sense similar to the
PNe.  c) (bottom right) {\em The Velocity Field for 99 Blue GCs.}  These 
are more extended than the red GCs, and have
kinematics less similar to the PNe.}
\end{figure}

\noindent In total, there are now 188 GCs
with measured velocities in Cen A, 125 of which are newly discovered
from our survey.

The smoothed PN velocity field shown in Figure~1
is the most detailed
and extensive stellar velocity field in an elliptical galaxy to date.
The rotation in the halo, discovered at 20 Kpc in previous work (H95), is 
now evident to the 80 Kpc limit of our survey.  Most
striking, however, is the previously undetected twist in the
zero-velocity contour.  Earlier work detected kinematic misalignments in
Cen~A and in other ellipticals (H95; Franx, Illingworth, \&
Heckman 1989), but this is some of the first compelling evidence for a
twist in the kinematic axis.  Misalignment and twisting of the kinematic
axis are predicted for triaxial galaxies seen in projection (Statler
1991), and also are seen in merger simulations (e.g. Bendo \& Barnes
2000). This program is an excellent advertisement for the potential of
future PNe surveys.  

Globular cluster candidates were identified from UBVRI CCD imaging based
on morphological and color criteria.  Radial velocities confirm
membership in the Cen A GC system ($V_{systemic} = 541$ km~s$^{-1}$), and provide
kinematic information. We constructed the velocity fields for the GCs
using our observations of new and known GCs, combined with 29 velocities
from Harris et al.\ (1992). The $V$--$I$ color distribution of GCs is
confirmed to be bimodal. Using the $V$--$I$ color, a proxy for metallicity in
old stellar systems, we can separate the GCs into blue (metal-poor) and
red (metal-rich) populations.  Figure 1 shows the spatial distribution
and velocity fields of the two GC populations.  The red GCs are more
centrally concentrated, and more rotationally supported than the blue
GCs.  The kinematics of the PNe are more like those of the red GCs than
the blue.  This is consistent with the field stars in Cen A being
predominantly metal-rich (Harris \& Harris 2001), and with the red
clusters and PNe having a common formation history. 

\acknowledgments

This research has made use of NASA's Astrophysics Data System Abstract Service and 
the NASA/IPAC Extragalactic Database (NED) which 
is operated by the Jet Propulsion Laboratory, California Institute of Technology,
under contract with the National Aeronautics and Space Administration. The authors
wish to thank the CTIO staff for their excellent help during the observing runs on
the CTIO 4-m telescope.  This research was supported in part by NASA grant NAG5-7697 
and
NSF grant AST-0098566.

\end{document}